\documentclass[runningheads]{llncs}
\usepackage[T1]{fontenc}
\usepackage{graphicx}
\usepackage{cite}
\usepackage{xspace}
\usepackage{listings}

\usepackage{float}
\usepackage{multirow}
\usepackage{mathpartir}
\usepackage{fancyvrb}
\usepackage{xcolor}
\usepackage{subcaption}
\usepackage{caption}
\usepackage{listings}
\usepackage{tikz}
\usepackage{syntax}
\usepackage{matlab-prettifier}
\usepackage{adjustbox}
\usepackage{xspace}
\usepackage{enumitem}
\usepackage{comment}
\usepackage{amsmath, bm}
\usepackage{wrapfig}
\usetikzlibrary{positioning}
\usepackage{stmaryrd}
\usetikzlibrary{calc}
\usetikzlibrary{fit}
\usepackage{relsize}
\usepackage{xfrac}
\usepackage{mathtools}
\usepackage{booktabs}

\definecolor{WowColor}{rgb}{.75,0,.75}
\definecolor{SubtleColor}{rgb}{0,0,.50}
\newcounter{margincounter}

\newcommand{\cf}{ConstraintFlow\xspace}

\newcommand{\var}{x}

\newcommand{\expr}{e}

\newcommand{\curr}{\cfkeywords{\texttt{curr}}\xspace}
\newcommand{\prev}{\cfkeywords{\texttt{prev}}\xspace}

\newcommand{\shape}{\cfkeywords{\texttt{shape}}\xspace}

\newcommand{\traverse}{\cfkeywords{\texttt{traverse}}\xspace}

\newcommand{\affine}{\cfkeywords{\texttt{Affine}}\xspace}
\newcommand{\relu}{\cfkeywords{\texttt{ReLU}}\xspace}
\newcommand{\maxpool}{\cfkeywords{\texttt{MaxPool}}\xspace}

\newcommand{\revrelu}{\cfkeywords{\texttt{rev\_ReLU}}\xspace}

\newcommand{\cfkeywords}[1]{\texttt{\footnotesize\bfseries\textcolor{keywords}{#1}}}
\newcommand{\cftypewords}[1]{\texttt{\footnotesize\bfseries\textcolor{typewords}{#1}}}

\newcommand{\defshape}{\cfkeywords{{\texttt{Def shape as}}}\xspace}

\newcommand{\polyexp}{\cftypewords{\texttt{PolyExp}}\xspace}

\newcommand{\ct}{\cftypewords{\texttt{Ct}}\xspace}
\newcommand{\float}{\cftypewords{\texttt{Real}}\xspace}
\newcommand{\intt}{\cftypewords{\texttt{Int}}\xspace}
\newcommand{\bool}{\cftypewords{\texttt{Bool}}\xspace}
\newcommand{\typeneuron}{\cftypewords{\texttt{Neuron}}\xspace}

\usepackage{graphicx}
\usepackage{caption}
\usepackage[skip=0.5ex]{subcaption}

\makeatletter
\newsavebox{\@brx}
\newcommand{\llangle}[1][]{\savebox{\@brx}{\(\m@th{#1\langle}\)}%
  \mathopen{\copy\@brx\kern-0.5\wd\@brx\usebox{\@brx}}}
\newcommand{\rrangle}[1][]{\savebox{\@brx}{\(\m@th{#1\rangle}\)}%
  \mathclose{\copy\@brx\kern-0.5\wd\@brx\usebox{\@brx}}}
\makeatother

\newcounter{number}

\definecolor{diagramcolor}{rgb}{0.70,0.0,0.56}
\definecolor{keywords}{rgb}{0.13,0.13,1}
\definecolor{typewords}{rgb}{0.1,0.6,0.2}
\definecolor{greencomments}{rgb}{0,0.5,0}
\definecolor{turqusnumbers}{rgb}{0.17,0.57,0.69}
\definecolor{redstrings}{rgb}{0.5,0,0}

\definecolor{codegreen}{rgb}{0,0.6,0}
\definecolor{codegray}{rgb}{0.5,0.5,0.5}
\definecolor{codepurple}{rgb}{0.58,0,0.82}
\definecolor{backcolour}{RGB}{250, 250, 250}

\lstdefinelanguage{ConstraintFlow}
    {morekeywords={def, shape, as, curr, prev, prev0, prev1, mapList, transformer, ReLU, Affine, Maxpool, DotProduct, rev_ReLU, rev_Affine, rev_Maxpool, rev_Max, rev_Min, rev_Add, rev_Mult, func, map, true, false, traverse, dot, flow, forward, backward, sum, layer, sym, compare, avg, len, max, min, and, in, solver, currList, equations, minimize, maximize, mult, add, sigmoid, tanh, max_u, max_l, argmax, sub},
    morekeywords = [4]{Bool, Int, Real, PolyExp, SymExp, Neuron, Noise, Ct},
    keywordstyle = \bfseries\color{keywords},
    keywordstyle = [4]{\bfseries\color{typewords}},
    sensitive=false, 
    morecomment=[l][\color{greencomments}]{///},
    morecomment=[l][\color{greencomments}]{//},
    morecomment=[s][\color{greencomments}]{{(*}{*)}},
    morestring=[b]",
    stringstyle=\color{redstrings}
    }

\lstnewenvironment{cflisting}
    {\lstset{language=ConstraintFlow,
                basicstyle=\ttfamily,
                breaklines=true,
                columns=fullflexible
    }}
    {}

\makeatletter
\lst@AddToHook{OnEmptyLine}{\addtocounter{lstnumber}{-1}}
\makeatother

\begin{document}
\title{ConstraintFlow: A DSL for Specification and Verification of Neural Network Analyses}
%
%
\author{Avaljot Singh\inst{1} \and
Yasmin Sarita\inst{1} \and
Charith Mendis\inst{1} \and 
Gagandeep Singh\inst{1,2}}
%
\institute{University of Illinois, Urbana Champaign \and 
VMware Research}
\maketitle              
\begin{abstract}
We develop a declarative DSL - \cf - that can be used to specify Abstract Interpretation-based DNN certifiers. In \cf, programmers can easily define various existing and new abstract domains and transformers, all within just a few 10s of Lines of Code as opposed to 1000s of LOCs of existing libraries. We provide lightweight automatic verification, which can be used to ensure the over-approximation-based soundness of the certifier code written in \cf for arbitrary (but bounded) DNN architectures. Using this automated verification procedure, for the first time, we can verify the soundness of state-of-the-art DNN certifiers for arbitrary DNN architectures, all within a few minutes. 
\end{abstract}

\section{Introduction}
Deep neural networks (DNNs) are the dominant AI technology currently and have shown impressive performance in diverse applications in diverse domains including computing, medicine, mathematics, and natural sciences.
However, their black-box construction~\cite{Ribeiro0G:16} and vulnerability against environmental and adversarial noise~\cite{lbfgs} have raised concerns about their trustworthiness in safety-critical settings such as autonomous driving and healthcare. The field of certified artificial intelligence (AI) to formally certify DNNs has emerged in recent years and has been proposed as a key technology to overcome this barrier in literature~\cite{nnvbook,robustmlsystems,trustworthyacm}


State-of-the-art DNN certifiers~\cite{deepz,refinezono,AI2,optAndAbs,deeppoly,zhang2018crown,alphacrown,star,cnncert,dutta,ehlers2017formal,scalablever,fastcrown,gpupoly,semidefinite,convexrelaxation,krelu,tjandraatmadja,vincent19,imagestars,wang2018neurify,wang2018,wang2021beta,weng18a,WongK18,wu2020,xiang2017,Zelazny2022OnOB,syrenn,fanc,ivan} are based on Abstract Interpretation~\cite{absi}. Developing a scalable certifier applicable to realistic networks requires balancing the tradeoff between certifier cost and precision. The current development  process is tedious and involves (i) coming up with algorithms that capture the intricate behaviors of DNNs by balancing the tradeoffs, and (ii) efficiently implementing them. This requires substantial expertise in algorithm design, formal methods, and optimizations. Even with the tedious development, existing certifier implementations~\cite{eran,Lirpa20,marabou} primarily concentrate on certifying properties for specific DNN topologies (e.g., feedforward) and are tailored to specific hardware architectures, lacking the ability to leverage machine-specific performance optimizations. 
However, as deep learning frameworks continually introduce new topologies, the manual development of custom implementations optimized for specific DNN topologies and hardware becomes increasingly cumbersome. Only a handful of experts have the required knowledge to develop a practical certifier which makes the development process inaccessible to most users of DNN technology.


In the context of DNN certification, the input to DNN is an infinite set of data points, usually specified as constraints over the possible values of the input layer neurons. These constraints act as the elements of an abstract domain. Instead of computing the values of the neurons, DNN certifiers propagate the constraints through the DNN by defining abstract transformers corresponding to each DNN operation. Certifier developers establish the \textit{soundness} of the certifiers by proving the soundness of the abstract transformers w.r.t the DNN operations. 


Currently, the DNN certifiers are implemented in large pieces of general-purpose programming languages such as Python, C++, etc. The commonly used libraries for DNN certification include auto\_LiRPA~\cite{Lirpa20}, ELINA~\cite{fastpoly}, and ERAN~\cite{deeppoly}. Further, the certifier developers usually prove the soundness of the ceritifiers by providing pen and paper proofs, which needs extensive mathematical background.

In this work, we propose a novel DSL - \cf -  to specify a minimal high-level description of DNN certifiers that includes specifying the tradeoff knobs and also the associated constraints. The tradeoff knobs can be used to adjust the scalability vs. precision tradeoffs of the certifier, while the specified constraints can be leveraged to automatically verify the soundness of the specified abstract transformers. 
The advantages of automatic verification, efficient implementation compilation, and compact certifier code in \cf facilitate wider adoption of DNN certification in contemporary AI development pipelines.
For example, Figure~\ref{fig:intro} shows that we can specify (and verify) the popular DeepPoly certifier in less than 15 lines of code (LOCs) in the new DSL as opposed to 1000s of LOCs in C++.

\begin{figure}
    \begin{lstlisting}
Def shape as (Real l, Real u, PolyExp L, PolyExp U) {[curr[l]<=curr, curr[u]>=curr, curr[L]<=curr, curr[U]>=curr]};

Func concretize_lower(Neuron n, Real c) = (c >= 0) ? (c * n[l]) : (c * n[u]);
Func concretize_upper(Neuron n, Real c) = (c >= 0) ? (c * n[u]) : (c * n[l]);

Func replace_lower(Neuron n, Real c) = (c >= 0) ? (c * n[L]) : (c * n[U]);
Func replace_upper(Neuron n, Real c) = (c >= 0) ? (c * n[U]) : (c * n[L]);

Func priority(Neuron n) = n[layer];

Func backsubs_lower(PolyExp e, Neuron n) = (e.traverse(backward, priority, false, replace_lower){e <= n}).map(concretize_lower);
Func backsubs_upper(PolyExp e, Neuron n) = (e.traverse(backward, priority, false, replace_upper){e >= n}).map(concretize_upper);

Transformer DeepPoly{
    Relu -> prev[l] > 0 ? (prev[l], prev[u], prev, prev) : (prev[u] < 0 ? (0, 0, 0, 0) : (0, prev[u], 0, ((prev[u] / (prev[u] - prev[l])) * prev) - ((prev[u] * prev[l]) / (prev[u] - prev[l]))));
    
    Affine -> (backsubs_lower(prev.dot(curr[w]) + curr[b], curr), backsubs_upper(prev.dot(curr[w]) + curr[b], curr), prev.dot(curr[w]) + curr[b], prev.dot(curr[w]) + curr[b]);
}

Flow(forward, -priority, false, DeepPoly);\end{lstlisting}
    \caption{\textbf{Verified} DeepPoly certifier specification in just 13 lines. This is a significant improvement over the existing implementation available at~\cite{ELINA} that contains thousands of lines of intricate C code.}
    \label{fig:intro} 
\end{figure}

\textbf{Main contributions}. 
\begin{itemize}[noitemsep, nolistsep]
    \item We present the preliminary research towards the development of \cf - DSL for specifying DNN certifiers that rely on Abstract Interpretation. 
    The users can define new transformers within existing domains and even create new abstract domains.
    \item We provide an automated bounded verification that can be used to check the soundness of the DNN certifier specification in \cf.
    \item We provide an extensive evaluation to show the practicality of \cf for specifying DNN certifiers and verifying them using the verification procedure.  
\end{itemize}

\section{DSL Design through Examples}
\subsection{Design Challenges.}
Abstract domains used in DNN certification often involve complex data structures, such as polyhedral expressions. In traditional implementations \cite{eran,Lirpa20}, neurons are represented as objects or structs that include their metadata and are connected according to the DNN architecture. This results in large arrays representing polyhedral expressions, utilizing pointers or similar constructs. Due to the intricacies involved in pointer arithmetic, it is hard to write code involving polyhedral expressions, while maintaining soundness.

For DNN certifiers, the input and output to the abstract transformers are abstract elements, which are conjunctions of constraints over all the neurons in the DNN. Given the large number of neurons and the complexity of each constraint, our DSL must succinctly represent the input and output.
Moreover, transformers can perform complicated operations over abstract elements like traversals through the graph that represents the DNN architecture. The DSL must provide constructs that enable such complicated operations while also being natural and concise. 
Further, DNN certification algorithms build different types of analysis using abstract domains, including forward analysis for computing the post-condition, backward analysis for computing the pre-condition, and their combinations. 
It is challenging to design a DSL that supports this wide variety of existing abstract domains, abstract transformers, and certification algorithms and is general enough to succinctly express  new practically useful designs.

\subsection{High-Level Approach}

In this preliminary DSL design for \cf, the abstract domain and transformers can be specified in functional form. The DSL design serves two purposes. First, the user specifies the \textit{constraints} associated with each neuron. These constraints are used by our automated verification procedure to verify the soundness of each abstract transformer individually in the certifier code. 
Second, for each DNN operation, the user can specify how the constraints \textit{flow} through the DNN.  
We now explain different parts of the DeepPoly specification in \cf shown in Fig.~\ref{fig:intro}.


\subsubsection{New Datatypes.}
Apart from the standard types including \float, \intt, and \bool, we identify the common datatypes and introduce them as types within the language. These include \typeneuron, \polyexp, \ct. All the neurons in the DNN are of the type \typeneuron. 
The type \polyexp represents the polyhedral expressions which are linear combinations over neurons. 
The constraints defined in the program are of the type $\ct$, for instance, $3 + 4n_1 \leq n_2$. 
Further, for any type $\gamma$, we also provide the corresponding list type which is represented as $\overline{\gamma}$. For example, the expression $\curr[w]$ is of the type $\overline{\float}$. Here, $w$ represents the learned parameters of the model associated with the \curr neuron when the \curr neuron is the result of an affine combination of neurons (e.g., a fully connected layer). 

%
Polyhedral expressions are commonly used in DNN certification algorithms, but handling them can be complicated for developers. We observe that much of the arithmetic involving them can be automated and so, we introduce them as first-class members, allowing access to operators like `+' which are overloaded in this case. 
To the best of our knowledge, no prior work has addressed the handling of polyhedral expressions at the language level.

\subsection{\textbf{Abstract Domain} }
In \cf, we focus on popular DNN certifiers that associate an abstract shape with each neuron separately, which imposes constraints on its possible values. 
The conjunction of the abstract shapes of all the neurons in the DNN forms an element of the abstract domain. So, to succinctly specify the abstract domain in \cf, we allow the user to specify the abstract shape associated with each neuron.  
For each DNN operation, users can write transformers that output an abstract shape associated with a single neuron - \curr (read as \textit{current}), and then the same transformer can be automatically applied to all neurons in the DNN, \curr acts as a syntactic placeholder for each neuron in the DNN. This approach offers a solution that reduces code size and complexity while facilitating easier verification of the transformer's soundness. 
The abstract shape is declared using the \shape construct -  \defshape $(\gamma_1 \ \var_1, \gamma_2 \ \var_2, \cdots) \{ \expr \}$, where $\gamma_i$ is a type, $\var_i$ is the name of a member of the abstract shape, and $\expr$ represents the constraints imposed on \curr by $\var_i$.
Using this construct, the DeepPoly shape and the associated constraints can be defined in \cf as shown in Line 1 of Figure~\ref{fig:intro}
, where, $l, u, L,$ and $U$ are the user-defined \textit{members} of abstract shape, which can be accessed using the square bracket notation (\curr[·]). 
$l, u$ are the concrete lower and upper bounds, and $L, U$ are the polyhedral lower and upper bounds of the neuron \curr. So, the constraints can be specified as $\{[\curr[l]<=\curr, \curr[u]>=\curr, \curr[L]<=\curr, \curr[U]>=\curr]\}$.


\subsection{\textbf{Abstract Transformers. }}
In this section, we briefly describe the \relu and \affine transformers defined in Fig.~\ref{fig:intro}. In Line 10, we show the DeepPoly abstract transformer for ReLU operation. Here, \prev is a singleton list of neurons containing the input to the ReLU operator and \curr is the output neuron. 
The code computes the convex approximation for \curr in 3 cases. First, when the input to the ReLU operator is completely in the positive region, i.e., $\prev[l] \geq 0$. In this case, the ReLU operator just acts as an identity function. So, the constraints for the neuron are the same as the constraints for its parent. That is, $\curr[l] = \prev[l]$, $\curr[u] = \prev[u]$, $\curr[L] = \prev[L]$, and $\curr[U] = \prev[U]$. In the \cf, if-then-else cases are described using the ternary operator $\_ ? \_ : \_$. 
In the second case, the input to the ReLu operator lies completely in the negative region. In this case, the output of the operator is always 0. So, in this case, $\curr[l] = \curr[u] = \curr[L] = \curr[U] = 0$.
In the third case, the concrete lower bound and the polyhedra lower bound of the ReLU operator are set to 0: $\curr[l] = \curr[L] = 0$, while the concrete upper bound is set to the concrete upper bound of the parent, $\curr[u] = \prev[u]$. The polyhedra upper bound of the operator is given by the equation: $\curr[U] = \frac{\prev[u]}{\prev[u] \ - \ \prev[l]} * \prev - \frac{\prev[u] \ * \ \prev[l]}{\prev[u] \ - \ \prev[l]}$


\subsubsection{Backsubstitution.} 
In the \cf code for the affine layer shown in Line 11 of Fig.~\ref{fig:intro}, the polyhedral bounds used in the DeepPoly algorithm are exact. For a neuron \curr in an affine layer, its lower and upper polyhedral bounds can be written as $L = U = \prev * \curr[w] + \curr[b]$ where $\prev$ is the list of input neurons of the \curr neuron, $w$ is the weight vector, and $b$ is the bias. 
However, as per the standard implementation~\cite{deeppoly}, calculating the interval lower and upper bounds for these constraints 
involves a backsubstitution step that computes an approximate solution to a linear program (LP) maximizing/minimizing the bounds for neuron \curr with respect to the DeepPoly constraints defined over neurons in all previous layers. This is the most expensive step of the DeepPoly analysis and involves traversing backward through the directed acyclic graph (DAG) describing the DNN architecture. At each step, it substitutes the neurons by their respective upper or lower bounds. In \cf, this substitution is specified by defining a neuron replacement function that given a neuron and its coefficient, returns an expression, either box or polyhedral, that it should be replaced with, in the input polyhedral expression. Replacement with box constraint is fast but also imprecise while the polyhedral replacement is more precise but has a higher cost.

\subsubsection{Tradeoffs.} 
It can be shown that the specific algorithm used for the graph traversal during backsubstitution can have an impact on the precision and cost of the algorithm. In particular, the Breadth-first search (BFS) can be shown to give better precision than Depth First Search (DFS). However, DFS provides more parallelization opportunities as compared to BFS. The exact algorithm for graph traversal can be specified by defining a priority function over the neurons in the polyhedral expression. This function associates a value to each of the neurons in the expression. The neuron with the highest priority is traversed next. The code for the affine approximation below uses a priority function specifying BFS traversal. Further, the precision and the time complexity of backsubstitution are also dependent on the depth to which the graph is traversed. A larger depth yields more precision but also increases cost. 
All these knobs can be changed to adjust the precision and the cost of the backsubstitution step. In \cf, we provide a construct for the graph traversal - $\traverse(f_1, f_2, f_3)$ where $f_1$ describes the termination criterion, $f_2$ describes the priority function and $f_3$ describes the neuron replacement function.
\section{Automated Soundness Verification} 
To ease the development of DNN certifiers, we to automatically verify the soundness of the certifier specification provided by the programmer. 

\subsubsection{DNN architectures.}
The input to the certifiers is a DNN, which is treated as a Directed Acyclic Graph (DAG) with neurons as the nodes connected by the DNN topology. The value of each neuron is the result of a DNN operation ($f$) whose input is the set of neurons that have outgoing edges leading to the given neuron. We refer to this set of neurons as \textit{previous} ($p$) in this discussion. There are various types of DNN operations ($f$) used in DNNs including primitive equations (Equation~\ref{eq:primitive}) like the Addition ($f_{add}$), Multiplication ($f_{mult}$), Maximum ($f_{max}$) of two neurons, ReLU of a neuron ($f_r$), Sigmoid of a neuron ($f_{\sigma}$), etc, and also composite operations (Equation~\ref{eq:affine}) that can be written as compositions of primitive operations, like Affine combination of neurons ($f_{a}$), MaxPool of a list of neurons ($f_m$), etc. In Equations~\ref{eq:primitive}, \ref{eq:affine}, the inputs to $f$ are the previous neurons represented as $p$. Note that the input to $f_r$ and $f_{\sigma}$ is a single neuron $p_1$, while that for $f_{add}, f_{mult}$ and $f_{max}$ are two neurons, namely $p_1$ and $p_2$. The input to $f_{a}$ and $f_m$ is a list of neurons represented as $\overline{p}$.  In the following equations, bias, and weights, $b, w_i$ are the DNN's learned parameters.
\begin{align}
\label{eq:primitive}
\begin{split}
    f_{add}(p_1, p_2) = p_1 + p_2  \qquad 
    f_{mult}(p_1, p_2) = p_1 &* p_2 \qquad 
    f_{max}(p_1, p_2) = \max(p_1, p_2) \\
    f_r(p) = \max(p, 0) \quad & \quad
    f_{\sigma}(p) = sigmoid(p)
\end{split}
\end{align}

\begin{equation}
\label{eq:affine}
    f_{a}(\overline{p}) = b + \sum_i w_i * \prev_i \qquad 
    f_{m}(\overline{p}) = \max_i(\prev_i)
\end{equation}

\subsubsection{SMT query.}
We encode the verification conditions as first-order logic formulae and feed them to an off-the-shelf SMT solver~\cite{z3}. The constructs provided in \cf are sufficient to naturally encode the various convex approximations (affine, ReLU, and Maxpool) and at the same time enable the automatic translation of the specifications to SMT queries.  
The verification of the ReLU approximation shown before is straightforward. We demonstrate a part of the SMT query generated for the more complicated Maxpool convex approximation applying max  over m neurons described below:
\begin{figure}
    \begin{lstlisting}[numbers=none, escapeinside={(*@}{@*)}]
Transformer DeepPoly(curr, prev){
    MaxPool -> max_l(prev) > max_u((sub(prev, argmax(prev, l)))) ?
                (max_l(prev), max_u(prev), argmax(prev, l), argmax(prev, l)) : 
                (max_l(prev), max_u(prev), argmax(prev, l), max_u(prev));
}
\end{lstlisting} 
        \caption{DeepPoly \maxpool Transformer}
        \label{fig:maxpool}
\end{figure}
Here, \texttt{max, len, avg} are inbuilt functions that have natural meanings that return the maximum value of a list, the length of a list, and the average value of a list respectively. 
The function \texttt{argmax(p, x)} returns the neuron with the maximum value of x from a list of neurons, p. The function \texttt{sub} is the set subtraction. 

In the following, we discuss the verification queries for the first branch of the maxpool specification. The query for the second branch is similar. For each path, the following query is generated.
$$\forall \ [\prev[l], \prev[u], \prev[L], \prev[U], \prev], \ \phi \implies \psi$$ where $\psi$ encodes the soundness verification conditions for the output neuron specified in Line 1 in Fig.~\ref{fig:intro} and $\phi = \phi_1 \wedge \phi_2 \wedge \phi_3$ where $\phi_1, \phi_2, \text{ and }\phi_3$ capture the conditions on input neuron list $\prev$, the path condition of the branch and the branch computation respectively.
$$\psi \ = \ (\curr[l] \leq \curr) \ \wedge \ (\curr[u] \geq \curr) \ \wedge \ (\curr[L] \leq \curr) \ \wedge \ (\curr[U] \geq \curr)$$
$$\phi_1 = (\prev[l] \leq \prev) \ \wedge \ (\prev[u] \geq \prev) \ \wedge \ (\prev[L] \leq \prev) \ \wedge \ (\prev[U] \geq \prev)$$
Formula $\phi^1_2 \wedge \phi^2_2$ captures the left hand side of the condition $l' = \max\_l(\prev)=\max(\prev_1[l], \ \prev_2[l], \cdots, \prev_m[l])$ where $\prev$ is a list of neurons, i.e., $\prev = [\prev_1, \prev_2, \cdots \prev_m]$.
$$\phi^1_2 \ = \bigvee_{i=1\cdots m} (l' = \prev_i[l]) ,\qquad \phi^2_2 \ = \bigwedge_{i=1\cdots m} (l' \geq \prev_i[l])$$ 
Similarly, the formula $\phi^3_2 \wedge \phi^4_2 \wedge \phi^5_2$ represents the condition $n = argmax(\prev, l)$.
$$\phi^3_2 \ = \bigvee_{i=1\cdots m} (n[l] = \prev_i[l]) ,\qquad \phi^4_2 \ = \bigwedge_{i=1\cdots m}(n[l] \geq \prev_i[l])$$
\begin{align*}
    \phi^5_2 \ = \bigwedge((n[l] = \prev_i[l]) & \implies\\
    (n[L] = \prev_i[L]) & \wedge (n[u] = \prev_i[u]) \wedge (n[U] = \prev_i[U])) \\
\end{align*}
Similarly, the formulae for \verb|sub| and \verb|max_u| are generated. So, $\phi_2$ captures the path condition.
$$\phi_2 = \phi^1_2 \wedge \phi^2_2 \wedge \phi^1_3 \wedge \phi^2_4 \wedge \phi_2^5 \cdots$$
Formula $\phi_3$ captures the exact implementation of the output of the first branch, i.e., $(\curr[l] = \max\_l(\prev))$, $(\curr[u] = \max\_u(\prev))$, $(\curr[L] = argmax(\prev, l))$ and $ (\curr[U] = argmax(\prev, l))$. The logic formula for this is also generated automatically by composing the encodings of the built-in constructs. We have also verified the DeepPoly affine approximation mentioned previously. However, it requires inductive reasoning and we omit it for brevity.

The verification for the \traverse construct is more challenging.  
Since we do not have any information about the DNN architecture, it is not possible to completely execute the loop specified in \traverse construct. 
%
Therefore, instead of unrolling the loop and using symbolic execution, we check the soundness of a user-provided invariant and then use this invariant to generate the verification query.
We check the soundness of this invariant in two steps. First, we verify that the invariant is satisfied at the beginning of the loop.
Second, we verify that for every input that satisfies the invariant, the output obtained by one iteration of the loop also satisfies the invariant. 
If the invariant is proved correct, we create a symbolic variable to represent that output of the expression involving the \traverse construct. We then assume that the invariant holds on this output. 

\color{black}

\section{Evaluation}
\label{sec:evaluation}
We implemented the automated verification procedure in Python and used Z3 SMT solver ~\cite{z3} for verification of different DNN certifiers. All of our experiments were run on a 2.50 GHz 16 core 11th Gen Intel i9-11900H CPU with a main memory of 64 GB.
In this section, we present the a preliminary evaluation and case studies to address the following research questions and to motivate the development of \cf.
\begin{enumerate}[start=1,label={\bfseries RQ\arabic*:}]
\item How easy is it to specify the \textbf{state-of-the-art DNN certifiers} in \cf? (Section~\ref{sec:rq1})
\item How easy is it to specify \textbf{new abstract shapes or transformers} in \cf? (Section~\ref{sec:newdefinitions})
\item Is it possible to \textbf{verify the soundness} of DNN certifiers in \cf? (Section~\ref{sec:rq3})
\item Is it possible to \textbf{detect unsound} abstract transformers in \cf? (Section~\ref{sec:rq4})
\end{enumerate}
\subsection{Writing existing DNN Certifiers in \cf}
\label{sec:rq1}
We investigated diverse state-of-the-art DNN certifiers based on Abstract Interpretation, including DeepPoly~\cite{deeppoly}, CROWN~\cite{zhang2018crown}, DeepZ~\cite{deepz}, RefineZono~\cite{refinezono}, Vegas~\cite{forwardbackward}, and IBP~\cite{interval} (Interval Bound Propagation). These widely used DNN certifiers cover a variety of abstract domains, transformers, and flow directions. 
In \cf, these certifiers can be specified declaratively, requiring substantially fewer Lines of Code (LOCs). 
For instance, in \cf, the abstract shape requires only 1 line, and the abstract transformers for these certifiers can be expressed in less than 45 LOCs. 
This contrasts sharply with existing frameworks~\cite{eran,Lirpa20}, where abstract shapes demand over 50 LOCs, and abstract transformers comprise 1000s of lines of intricate, unverified code in general-purpose programming languages.
Notably, we can also handle various flow directions effectively. For instance, the Vegas certifier~\cite{forwardbackward}, which employs both forward and backward flows, is easily expressed in \cf.
The transformer for the forward direction is the same as the DeepPoly analysis, while the transformer for the backward analysis replaces operations like $\relu$ with $\revrelu$.
    
    

\subsection{Defining new Abstract Shapes and new Abstract Transformers}
\label{sec:newdefinitions}
\subsubsection{New Abstract Shapes. }
\cf is powerful enough to allow users to easily design new abstract domains. 
Just by listing the members associated with a neuron and the corresponding constraints that are imposed on neuron's possible values, one can easily define a new abstract shape. Since, we define specialized datatypes for commonly used data structures in DNN certification, coming up with novel abstract shapes is easy. For example, one can maintain two polyhedral lower and upper bounds instead of one to write more precise abstract domains:
\begin{lstlisting}[numbers=none]
def shape as (Real l, Real u, PolyExp L1, PolyExp U1, PolyExp L2, PolyExp U2) {curr[l] <= curr and curr[u] >= curr and curr[L1] <= curr and curr[U1] >= curr and curr[L2] <= curr and curr[U2] >= curr};  
\end{lstlisting}
This can be particularly helpful in certifiers like DeepPoly, where there are multiple options for choosing appropriate lower and upper polyhedral bounds. Depending on the downstream application, the users can define the polyhedral bounds to adjust the scalability-precision tradeoffs.

\subsubsection{New Abstract Transformers. }
\label{sec:newcert}
\begin{figure}
    \begin{lstlisting}[numbers=none, escapeinside={(*@}{@*)}]
Transformer DeepPoly(curr, prev){
    MaxPool -> max_l(prev) (*@\textcolor{red}{\textbf{>=}}@*) max_u((sub(prev, argmax(prev, l)))) ?
                (max_l(prev), max_u(prev), argmax(prev, l), argmax(prev, l)) : 
                (max_l(prev), max_u(prev), argmax(prev, l), max_u(prev));
}
\end{lstlisting} 
        \caption{New DeepPoly \maxpool Transformer}
        \label{fig:newmaxpool}
\end{figure}
We can also concisely specify new abstract transformers for DNN certifiers. For instance, we designed a new  \maxpool abstract transformer for the DeepPoly certifier. 
The standard DeepPoly \maxpool transformer checks if there is a neuron in \prev whose lower bound is greater than the upper bounds of all other neurons, i.e., $\exists p \in \prev \cdot \forall q \in (\prev\setminus\{p\}) \cdot p[l] \color{red}\bm{>} \color{black}q[u]$. If such a neuron $p$ exists, then that neuron is the output of the \maxpool operation. Otherwise, the lower and upper polyhedral bounds are equal to the lower and upper concrete bounds. 
However, there can be cases where a neuron's lower bound is equal to the maximum of the upper bounds of other neurons. For example, let the neurons in \prev be $n_1, n_2, n_3$, s.t., $n_1 \in [0, 1], n_2 \in [1, 2], n_3 \in [2, 3]$. Here, the polyhedral lower bound for \maxpool transformer using DeepPoly's standard implementation is $\max(0, 1, 2) = 2$. In such cases, we can prove that setting the lower polyhedral bound to $n_3$ is sound using the our verification procedure. Further, it 
has been shown previously~\cite{deeppoly} that polyhedral bounds are more precise than interval bounds. So, if $\exists p \in \prev \cdot \forall q \in (\prev\setminus\{p\}) \cdot p[l] \color{red}\bm{\geq} \color{black} q[u]$, then it is more precise to set the lower bound to $p$. 
%
The code for the new \maxpool transformer is shown in Figure~\ref{fig:newmaxpool}. 


\subsection{Verifying Soundness of DNN Certifiers}
\label{sec:rq3}
Using \cf's bounded automatic verification, for the first time, we can prove the over-approximation-based soundness of the popular DNN certifiers for arbitrary (but bounded) DNNs. 
%
In \cf, although verifying transformers for primitive operations directly implies the verification of arbitrary compositions, in some cases, transformers can be more precise if specified directly for composite operations. So, in Tables~\ref{table:primitivecertifiers} and ~\ref{table:complexcertifiers}, we evaluate transformers for primitive and composite separately and present the time taken to generate the query (Column G) and the time taken by the SMT solver to prove the query (Column V). 
The primitive operations - \relu, \cfkeywords{Max}, \cfkeywords{Min}, \cfkeywords{Add} shown in Table~\ref{table:primitivecertifiers} can be verified in fractions of a second for all existing certifiers. Note that no bounds are assumed for the verification of primitive operations.
For verification of composite operations - \affine and \maxpool - shown in Table~\ref{table:complexcertifiers}, the parameters, $n_{prev}$ (maximum number of neurons in a layer) and $n_{sym}$ (maximum length of a polyhedral or symbolic expression) are used during the graph expansion step and impact the verification times. For our experiments, we set $n_{sym} = n_{prev}$. 
Note that $n_{prev}$ is an upper bound for the maximum number of neurons in a single affine layer, and does not impose any restriction on the total neuron count in the DNN. So, the DNN can have an arbitrary number of affine layers, each with at most $n_{prev}$ neurons, allowing for an arbitrary total number of neurons.

\textbf{Parameter values. }
We determine these parameter values through an analysis of DNNs that existing DNN certifiers can handle~\cite{gpupoly,deeppoly,zhang2018crown,cloud}. These can handle a maximum of $10$ neurons for \maxpool. The transformer for \affine includes DNN operations like convolution layers, average pooling layers, and fully connected layers. Typical convolution and average pooling kernel sizes that are handled by existing DNN certifiers are $3 \times 3 \times 3, 64 \times 3 \times 3, 128 \times 3 \times 3$, etc., resulting in $n_{prev} = 9, 576, 1152$. For fully connected layers, the most commonly used values for $n_{prev}$ are $512, 1024, 2048$. So, in Table~\ref{table:complexcertifiers}, we present the computation times for \affine with $n_{prev}=2048$ and \maxpool transformers with $n_{prev}=10$. 

\begin{table}
\centering
\captionsetup{justification=centering}
\resizebox{\textwidth}{!}{
\setlength{\tabcolsep}{5pt} 
\begin{tabular}{@{}l | c  c c | c c c | c  c c | c c c @{}}
\toprule
Certifiers & \multicolumn{3}{c}{\relu} & \multicolumn{3}{c}{\cfkeywords{Max}} & \multicolumn{3}{c}{\cfkeywords{Min}} & \multicolumn{3}{c}{\cfkeywords{Add}}\\
\text{ } & G & V  & B & G & V  & B & G & V  & B & G & V  & B \\ 
\hline
DeepPoly & 0.15 & 0.96 & 1.22 & 0.08 & 1.69 & 0.18 & 0.07 & 1.84 & 0.19 & 0.06 & 0.09 & 0.27 \\ 
Vegas & 0.12 & 0.37 & 0.32 & 0.04 & 0.07 & 0.17 & 0.04 & 0.07 & 0.25 & 0.04 & 0.05 & 0.21   \\ 
DeepZ & 0.11 & 0.48 & 0.48 & 0.13 & 0.66 & 0.17 & 0.13 & 0.65 & 0.19 & 0.07  & 0.06 & 0.06  \\ 
RefineZono & 0.13 & 0.52 & 0.53 &  0.14 & 0.66 & 0.87 & 0.14 & 0.69 & 0.14 & 0.07 & 0.07 & 0.23  \\ 
IBP & 0.09 & 0.31 & 0.14 & 0.11 & 0.32 & 0.15 & 0.11 & 0.32 & 0.12 & 0.06 & 0.04 & 0.16  \\ 
\midrule
\end{tabular}
}
\caption{Query generation time (G), verification time (V) for correct implementation, and bug-finding time for randomly introduced bugs (B) in seconds for transformers of primitive operations.}
\label{table:primitivecertifiers}
\end{table}

\begin{table}
\centering
\captionsetup{justification=centering}
\setlength{\tabcolsep}{6pt} 
\begin{tabular}{@{}l | c  c  c | c  c  c@{}}
\toprule
\multirow{2}{*}{Certifiers} & \multicolumn{3}{c}{\affine} & \multicolumn{3}{c}{\maxpool} \\ 
& G & V  & B & G & V  & B   \\ 
\hline
DeepPoly  &  18.43 & 1772.83 & 455.27 & 15.34 & 369.60 &  81.18 \\ 
Vegas  & 2.17 & 8.83 & 4.39 & - & - & -\\ 
DeepZ & 16.32 & 898.76 & 12.21 & 12.43 & 444.14  & 406.96 \\ 
RefineZono & 5.07 & 353.09 & 4.40 & 12.16 & 400.43 & 360.72 \\ 
IBP & 14.78 & 590.61 & 34.13 & 0.09 & 4.35 & 0.08 \\ 
\midrule
\end{tabular}
\captionof{table}{Query generation time (G), verification time (V) for correct implementation, and bug-finding time for randomly introduced bugs (B) in s for transformers of composite operations.}
\label{table:complexcertifiers}
\end{table}

\subsection{Detecting Unsound Abstract Transformers}
\label{sec:rq4}
Using \cf, we can identify unsound abstract transformers. 
We randomly introduce bugs in the existing DNN certifiers. The bugs were introduced for all DNN operations and we were able to detect unsoundness in all of them. 
These include (i) changing the operations to other operations with similar types of operands, e.g., + to -, max to min, etc., (ii) changing the shape member to another shape member with the same type, e.g., curr[l] to curr[u], (iii) changing function calls to other functions with the same signature, and (iv) changing the neurons, e.g., \prev to \curr, when \prev represents a single neuron.
%
The bug-finding times for the incorrect implementations of primitive operations are shown in Table~\ref{table:primitivecertifiers} - column B, while those for composite operations are shown in Table~\ref{table:complexcertifiers} - column B. For \affine, we use $n_{sym} = n_{prev}=2048$, and for \maxpool, $n_{sym} = n_{prev}=10$. 
Similar to the verification times for correct implementations, the bug-finding times are usually less than a second for primitive operations.  
For composite operations, the bug-finding times are always less than the verification times because to disprove the query it is sufficient to find one counter-example.
%

\section{Related Work}
\label{sec:related}
\noindent\textbf{DNN Certification. }
The recent advancements in DNN certification techniques~\cite{albarghouthi} have led to the organization of competitions to showcase DNN certification capabilities~\cite{vnncomp}, the creation of benchmark datasets~\cite{vnnlib}, the introduction of a DSL for specifying certification properties~\cite{reliableneuralspecification,dnnv}, and the development of a library for DNN certifiers~\cite{li2023sok,socrates}. 
However, they lack formal soundness guarantees and do not offer a systematic approach to designing new certifiers.

\noindent\textbf{Symbolic Execution. }
To handle external function calls, ~\cite{dart,cute,exe} use concrete arguments creating unsoundness. To improve on this, ~\cite{klee,s2e,cloud9} create an abstract model to capture the semantics of the external calls. In the context of DNN certification, the external calls are restricted to optimizers such as LP, MILP, and SMT solvers. 
This restriction makes it possible to simulate their behavior without actually executing them on concrete inputs or creating extensive abstract models.



Invariant synthesis is done by various techniques including computing interpolants~\cite{interpolants}, using widening operators and weakest preconditions for path subsumption~\cite{tracer,abstractloops,abstractloops2}, and
learning-based techniques~\cite{mayur,DBLP:journals/corr/abs-1712-09418,DBLP:conf/cav/0001LMN14,DBLP:journals/pacmpl/EzudheenND0M18}. 
These methods suffer from problems like long termination times, generating insufficient invariants, etc. 
However, if the property to be proven is known beforehand, it can be used to guide the search for a suitable invariant. Fortunately, in \cf, 
the soundness property is known beforehand. This gives an opportunity to automatically synthesize an invariant for \traverse in the future. 
\noindent\textbf{Verification of the Certifier.}
The correctness of certifiers can be verified using Abstract Interpretation~\cite{a2i}, or by translating the verification problem to first-order logic queries and using off-the-shelf solvers~\cite{ebpf,10.1145/3586029}. Some existing works prove the correctness of the symbolic execution w.r.t the language semantics~\cite{10.1145/3547628}. However, these methods establish correctness in the scenario where symbolic execution also represents symbolic variables used in concrete executions.  
\section{Conclusion and Future Work}
\label{sec:discussion}
We presented a preliminary version of a new DSL - \cf - to specify DNN certifiers in a pointer-free, declarative manner. In \cf, we aim to provide new datatypes and constructs to easily write various existing and new certifiers by decoupling their specification from implementation. 
We introduce a procedure enabling for the first time, the verification of the soundness of state-of-the-art DNN certifiers for arbitrary DAG topologies. In the future, we aim to develop \cf to address the growing concerns about AI safety and trustworthiness and make DNN certification accessible, scalable, and easy to use and develop. 
\subsubsection{Formal Semantics.} 
We aim to develop formal semantics for the constructs introduced in \cf. By studying state-of-the-art DNN certifiers, commonalities can be identified, which can then be made first-class members of the language design. The formal semantics would provide the foundation for the development of more complex static and dynamic analyses to check the soundness of the specifications.   

\subsubsection{Compiler.} 
We aim to develop a compiler framework that can generate precise, scalable, fast, and memory-efficient code for certifying properties for arbitrary DNNs. The framework will optimize the code for different hardware based on a high-level description of certifier logic.
Note that, in contrast to existing tensor compilers that propagate tensor values and optimize mainly for runtime, our compiler will determine how \emph{constraints} will be propagated to balance the precision, speed, and memory tradeoffs. We plan to introduce novel compiler transformations to explore the optimization space and propose a new auto-tuning system to automatically traverse it.

\subsubsection{Automatic Synthesis of Abstract Interpreters.} 
Recent work on automating abstract interpreters highlights the need for tools that can automate search-space exploration, construct sound transformers, and optimize precision. ConstraintFlow addresses these challenges directly: its DSL can be used to specify the abstract domain and the transformers, the semantics of the constructs define the DSL's behavior, and its automated verification ensures soundness. This makes ConstraintFlow a powerful platform for developing and refining abstract interpreters. The users can thus develop algorithms to automatically generate sound Abstract Interpretation-based DNN certifiers in \cf.

Further, the preliminary design of \cf supports customized polyhedral expressions extensively used for  DNN certification. However, we believe that \cf design can be extended for general program analysis by supporting other specialized domains like Octagon and Pentagon~\cite{octagon,pentagon,10.1145/2813885.2738000}.

\subsubsection{Non-linear Activation Functions.}
The verification procedure currently does not support non-linear activations like Sigmoid, Tanh, etc. This is because it is limited by the unpredictable behavior and the exponential time complexity of the SMT solvers, and also the undecidability of the non-linear operations~\cite{DBLP:journals/corr/abs-2305-06064}. The verification procedure can still be soundly extended to verify non-linear operations by creating their linear over-approximations. 
\clearpage
\bibliographystyle{splncs04}
\bibliography{main}

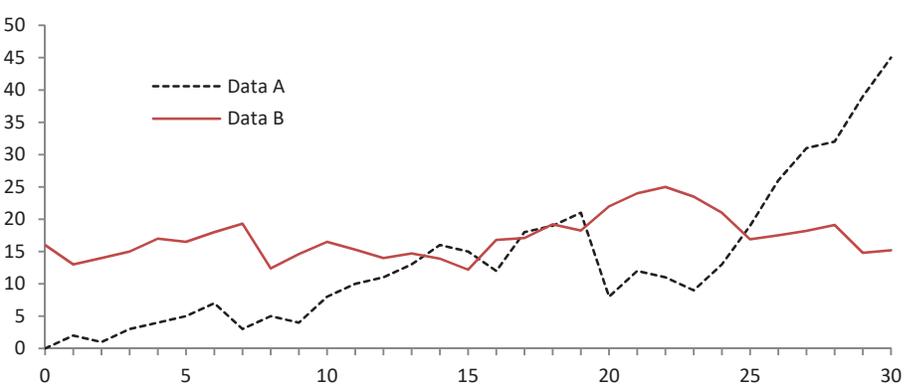
\end{document}